\documentclass[aps,prb,twocolumn]{revtex4-1}

\usepackage{amsmath}
\usepackage{amssymb}

\usepackage{xcolor}

\usepackage{setspace}
\usepackage{bm}
\usepackage{graphicx}
\numberwithin{equation}{section}
\usepackage{hyperref}
\usepackage{subfig}
\usepackage{braket} 

\newcommand{\udubee}{Department of Electrical Engineering, University of Washington, Seattle, WA 98195, USA}
\newcommand{\udubphys}{Department of Physics, University of Washington, Seattle, WA 98195, USA}
\newcommand{\iit}{Department of Electrical Engineering, Indian Institute of Technology, Delhi, Hauz Khas, New Delhi 110016, India}

\begin{document}
\title{Strong photon antibunching in weakly nonlinear two-dimensional exciton-polaritons}
\author{Albert Ryou$^1$, David Rosser$^2$, Abhi Saxena$^3$, Taylor Fryett$^1$, Arka Majumdar$^{1,2}$}
\affiliation{
$^1$ \udubee \\
$^2$ \udubphys \\
$^3$ \iit}
\date{\today}

\begin{abstract}
A deterministic and scalable array of single photon nonlinearities in the solid state holds great potential for both fundamental physics and technological applications, but its realization has proved extremely challenging. Despite significant advances, leading candidates such as quantum dots and group III-V quantum wells have yet to overcome their respective bottlenecks in random positioning and weak nonlinearity. Here we consider a hybrid light-matter platform, marrying an atomically thin two-dimensional material to a photonic crystal cavity, and analyze its second-order coherence function. We identify several mechanisms for photon antibunching under different system parameters, including one characterized by large dissipation and weak nonlinearity. Finally, we show that by patterning the two-dimensional material into different sizes, we can drive our system dynamics from a coherent state into a regime of strong antibunching with $g^{(2)}(0) \sim 10^{-3}$, opening a possible route to building scalable, on-chip quantum simulators.
\end{abstract}

\maketitle


\begin{figure}
\includegraphics[width=83mm]{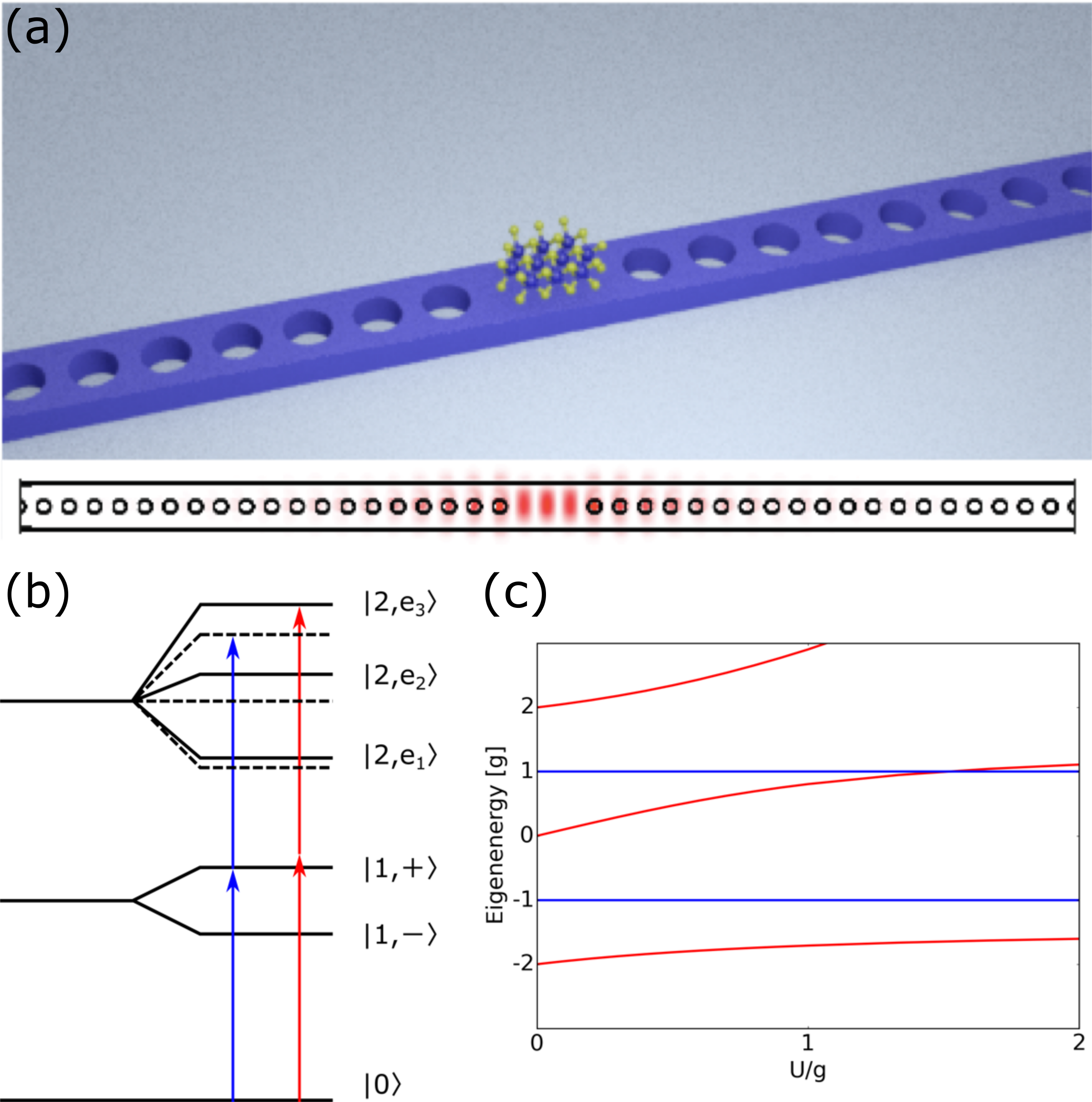}
\caption{\label{Figure:cavity_cartoon} (Color online) \textbf{Patterned 2D material-embedded cavity}. (a) Schematic illustration of the proposed experimental platform. A patterned 2D-material (tungsten diselenide, WSe$_2$) monolayer is placed on top of a photonic crystal nanobeam cavity. The radius of the monolayer is on the order of tens of nanometers. The top view of the cavity with a simulated field profile of the fundamental mode is shown below. The calculated mode volume is about $2.5(\lambda/n)^3$. (b) Energy level diagram. The dressed states are labeled by the number of energy quanta, or Fock manifold, followed by a symbol: $\ket{1, -}$ and $\ket{1,+}$ are the first-manifold states representing the lower and upper polaritons; $\ket{2, e_1}$, $\ket{2, e_2}$, and $\ket{2, e_3}$ are the second-manifold states. The solid lines represent the eigenenergies of the Hamiltonian with nonzero nonlinearity, whereas the dotted lines represent the eigenenergies with zero nonlinearity. The arrows represent the pump laser frequency that is resonant with either $\ket{1,+}$ (blue) or $\ket{2, e_3}$ (red). (c) Eigenenergies as a function of the nonlinearity U, calculated via exact matrix diagonalization. All parameters are normalized by the exciton-photon coupling strength $g$.} 
\end{figure}

\section{\label{sec:Intro} Introduction}

Quantum optical nonlinearities have received growing interest for their key role in quantum information science \cite{kiraz2004quantum}, quantum simulations \cite{kuhn2002deterministic}, and other quantum technologies \cite{milburn1989quantum}. While nonlinear effects with individual emitters have been demonstrated across a range of platforms, including ultracold atoms \cite{bakr2009quantum}, superconducting qubits \cite{devoret2013superconducting}, and semiconductor quantum dots \cite{Photon_Blockade_AM, Photon_Blockade_AF}, realizing a deterministic and scalable array of such nonlinearities has proved a far more challenging task. For quantum dots, which are particularly attractive due to their versatility and on-chip compatibility \cite{Photon_Blockade_Atac}, random positioning and inhomogeneous broadening of the emitters remain formidable bottlenecks \cite{QW_strain, Quantum_simulation_CCA_Hartman}.

Another solid-state candidate for quantum nonlinear optics is the exciton-polariton, a quasiparticle made of a semiconductor exciton strongly coupled to a microcavity photon. Inheriting strong interactions from the matter component and fast dynamics and state observability from the photonic component, exciton-polaritons are particularly well-suited as building blocks for photonic quantum simulations \cite{Na_Young_Kim_EP_Simulator, Quantum_fluids_light, Fermionized_Photons}. A host of many-body correlated phenomena with exciton-polaritons have been observed, including Bose-Einstein condensation \cite{Hui_deng_EP_BEC} and polariton lasing \cite{Polariton_devices}. Nevertheless, there has been no report of a strong polariton-polariton interaction at a single quantum level. To increase the interaction strength, several researchers tried shrinking the size of the polariton wavefunction. Besga \textit{et al.} decreased the cavity mode volume by employing a fiber-tip cavity \cite{besga2015polariton}, and recently Mu{\~n}oz-Matutano \textit{et al.}, using a similar setup, reported a weak nonlinearity \cite{munoz2017quantum}. Researchers have also tried decreasing the effective size of group III-V quantum wells, albeit with limited success \cite{verma2008high, lee2011room}.

\begin{figure*}
\includegraphics[width=175mm]{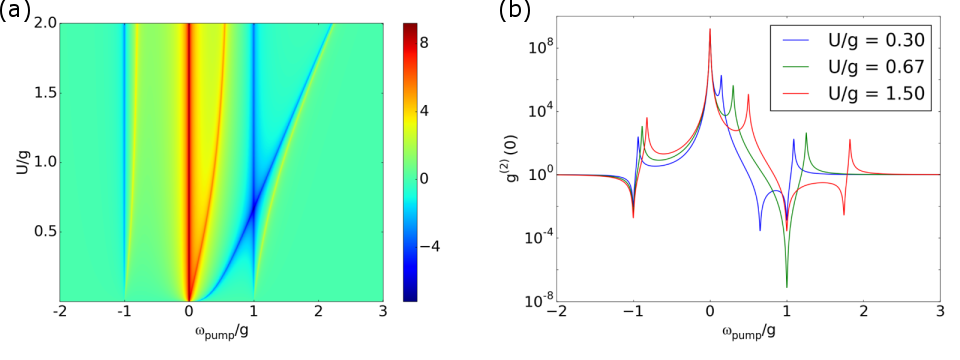}
\caption{\label{Figure:g2_vs_detuning_vs_U} (Color online) \textbf{$g^{(2)}(0)$ vs. pump laser frequency for different U}. (a) A 2D plot of $g^{(2)}(0)$ versus pump laser frequency detuning (x-axis) for different values of U (y-axis). The color corresponds to the base-10 logarithm of $g^{(2)}(0)$. Four strong bunching peaks (red) are observed, three of which come from the second-manifold eigenstates. The remaining bunching peak at $\omega_{pump} = 0$ is due to photon-induced tunneling \cite{Photon_Blockade_AF}. Also observed are three strong antibunching dips (blue): the first-manifold eigenstates (lower and upper polaritons) and a quantum-interference dip. The other parameters are $\omega_e = \omega_c = 0$ and $\kappa = \Gamma = 0.01g$. (b) Horizontal cross-sections of (a) for U/$g$ = 0.3, 0.67, and 1.5. When U/$g$ is near 2/3, the location of the quantum interference dip overlaps with that of the upper polariton at $\omega_{pump} = g$, yielding an extremely strong antibunching with $g^{(2)}(0) \sim 10^{-7}$.}
\end{figure*}

Recent advances in atomically thin two-dimensional (2D) materials point to a new potential platform for scalable quantum optical nonlinearities. These materials, including graphene, hexagonal boron nitride, and transition-metal dichalcogenides (TMDCs), boast exceptional light-emitting and light-harvesting properties, along with an unprecedented ability to be fabricated and transferred onto other photonic structures \cite{fryett2017cavity}. The TMDCs, in particular, hold great promise for integrated photonics due to their large, direct bandgap \cite{TMDC_optoelectronics}. TMDCs embedded in microcavities have been employed to observe optically pumped lasing \cite{Laser_AM_Xu, Laser_Xiang}, cavity-enhanced electroluminescence \cite{CHL_LED}, second harmonic generation \cite{TKF_SHG}, and strong coupling \cite{Vinod_Menon_EP, Sasha_EP}. Finally, Wei \textit{et al.} showed that TMDCs patterned via electron beam lithography into circular nanodots with radii down to 15 nm could still host long-lived excitons \cite{wei2017size}.

In this paper, we analyze the optical nonlinearity of a 2D-material monolayer coupled to a low mode-volume photonic crystal defect cavity. The strength of the quantum interaction can be revealed by its second-order coherence function $g^{(2)}(\tau)$. We identify different mechanisms that give rise to non-classical photon distributions and arrive at a robust regime, characterized by large dissipation and weak nonlinearity, whose second-order coherence at zero time delay is much less than unity. Finally, we consider the effect of the size of the monolayer on the system parameters. We numerically show that by physically patterning the monolayer into different sizes, it is possible to drive its dynamics from a coherent state into a non-classical regime with $g^{(2)}(0)\sim 10^{-3}$. An observation of such strong photon antibunching in this hybrid platform would open the door to further experiments in coupled nonlinear cavities and scalable quantum simulators.


\section{\label{sec:Para1} System description}

Our system consists of a patterned 2D-material monolayer placed on top of a photonic crystal nanobeam cavity (see Fig. 1a) \cite{fryett2017encapsulated}. The choice of a nanobeam has been motivated by its small cavity mode volume. The simulated field profile of the fundamental mode of the cavity is shown below the schematic. Unlike the conventional semiconductor-embedded distributed Bragg reflector cavity, whose excitons couple to a continuum of in-plane momenta, the monolayer-embedded photonic crystal cavity only supports a narrow band in the momentum space. Thus, in our model we consider only those excitons whose momenta match that of the fundamental cavity mode \cite{munoz2017quantum}.

In a frame rotating at the frequency of an external pump laser, the Hamiltonian of a strongly coupled exciton-polariton system is given by (setting $\hbar = 1$)    
\begin{equation}
H = \Delta_c a^\dagger a + \Delta_e b^\dagger b + g \left(a^\dagger b + a b^\dagger \right) + Ub^\dagger b^\dagger b b + E(a^\dagger + a)
\end{equation}
where $a^\dagger (a)$ and $b^\dagger (b)$ are the creation (annihilation) operators for the cavity photon and the monolayer exciton, respectively; $\Delta_c = \omega_c - \omega_{pump}$ and $\Delta_e = \omega_e - \omega_{pump}$ are their frequency detunings relative to the pump laser; $g$ is the exciton-photon coupling strength; U is the on-site Kerr nonlinearity representing the exciton-exciton repulsion \cite{Sun2017}; and E is the strength of the pump laser. The system dynamics is given by the evolution of the density matrix according to the master equation \cite{kavokin2007microcavities}:
\begin{align}
i \dot{\rho} = &\left[H,\rho\right] + i\frac{\kappa}{2} \left( 2a\rho a^\dagger - a^\dagger a \rho - \rho a^\dagger a \right) \nonumber \\
&+ i\frac{\Gamma}{2} \left( 2b\rho b^\dagger - b^\dagger b \rho - \rho b^\dagger b \right)
\end{align}
where $\kappa$ and $\Gamma$ are the inverse lifetimes of the cavity photon and the exciton, respectively.

\begin{figure*}
\includegraphics[width=175mm,trim=20 5 5 15,clip]{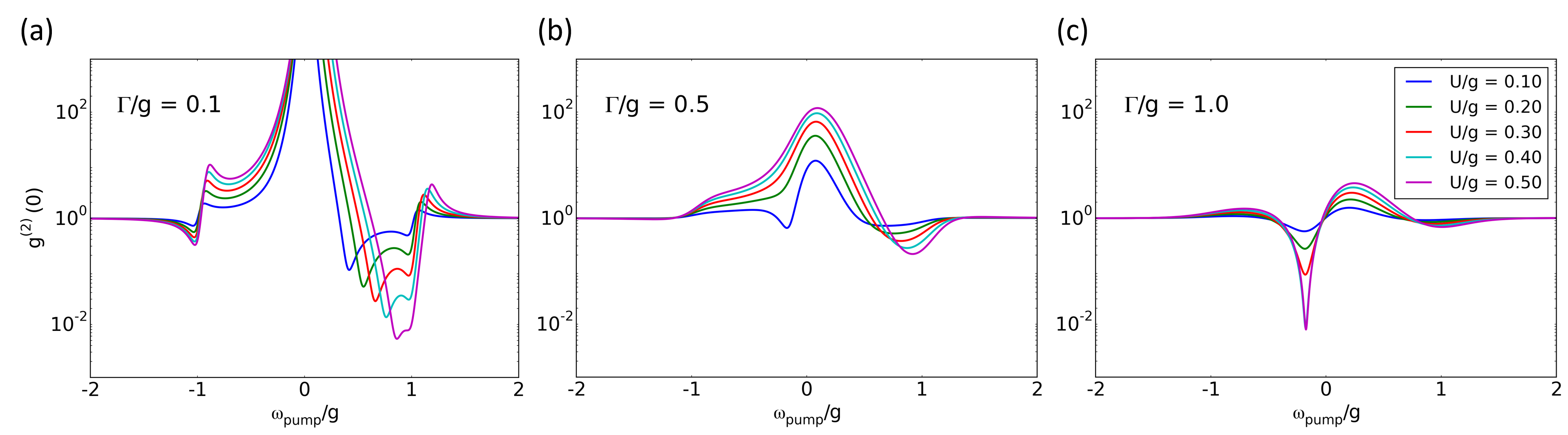}
\caption{\label{Figure:g2_vs_detuning_vs_multiple_U} (Color online) \textbf{$g^{(2)}(0)$ vs pump laser frequency for different $\Gamma$ and U}. (a-c) $\Gamma$/$g$ = 0.1, 0.5, and 1.0, with U/$g$ ranging from 0.1 to 0.5. The pump laser frequency is relative to the exciton resonance, and $\omega_e = \omega_c = 0$. (a) For small $\Gamma$, $g^{(2)}(0)$ resembles that in Fig. 2b, with the strong quantum interference-induced antibunching appearing near $\omega_{pump} = g$. (b) For intermediate $\Gamma$, the antibunching dip at $\omega_{pump} = g$ becomes shallow while a new antibunching dip appears at a slightly negative $\omega_{pump}$. (c) This new antibunching dip, also due the destructive quantum interference, can be significantly large with $g^{(2)}(0) \sim 10^{-2}$.}
\end{figure*}

The energy level diagram of the system containing up to two energy quanta is shown in Fig. 1b, where we have taken $\omega_c = \omega_e$. The degeneracy of the bare states is lifted by the exciton-photon coupling. The dressed states $\ket{1, -}$ and $\ket{1,+}$, containing one energy quantum and collectively known as the first Fock manifold of the Hamiltonian, represent the lower and upper polaritons, respectively. Similarly, the second-manifold states, $\ket{2, e_1}$, $\ket{2,e_2}$, $\ket{2,e_3}$, containing two energy quanta, become nondegenerate. For zero exciton-exciton repulsion (U = 0), their eigenenergies are $-2g$, 0, and $2g$ (dotted lines), forming a harmonic energy ladder for two coupled oscillators. For U $>0$, however, the eigenenergies shift (solid lines). The eigenenergies of the first (blue) and the second (red) manifold as a function of U are plotted in the rotating frame in Fig. 1c.

The shifting of the second-manifold eigenenergies due to the exciton-exciton repulsion is the source of the quantum optical nonlinearity. Consider tuning the pump laser so that it resonantly excites the upper polariton $\ket{1, +}$ (blue arrows in Fig. 1b). Whereas the first photon from the laser drives the system from $\ket{0}$ to $\ket{1, +}$, a second photon cannot subsequently drive the system from $\ket{1, +}$ to $\ket{2, e_3}$ because the eigenenergy of $\ket{2, e_3}$ has shifted out of resonance. On the other hand, if the pump laser is tuned to half the energy of $\ket{2, e_3}$ (red arrows), it can no longer excite $\ket{1, +}$, while at the same time, it can excite $\ket{2, e_3}$ via two-photon resonance. Thus, by measuring the photonic content of the state of the system, we can determine the strength of the nonlinearity.

The photonic content, in turn, can be measured by detecting the light that leaks out of the cavity and analyzing its temporal distribution. The second-order coherence function $g^{(2)}(\tau)$ yields the ratio of the detection rate of photon pairs separated by a delay $\tau$ to that of single photons: 
\begin{equation}
g^{(2)}(\tau) = \frac{\braket{a^\dagger (0) a^\dagger(\tau) a(\tau) a(0)}}{\braket{a^\dagger(0) a(0)}^2}
\end{equation}
In particular, for zero time delay, $g^{(2)}(0) = 1$ indicates a Poissonian distribution typical of classical light, whereas $g^{(2)}(0) < 1$ is a sub-Poissonian distribution and an experimental smoking gun of a distinctly quantum process. In the following section, we will investigate $g^{(2)}(0)$ in various parameter spaces.


\section{\label{sec:Para1} Parameter study of $g^{(2)}(0)$}

We first consider $g^{(2)}(0)$ for $\Gamma \ll g \ll $ U. We assume $\kappa$ is equal to $\Gamma$. The second-manifold eigenenergies approach $\pm \sqrt{2}g$ and 2U, the former pair resembling the well-known anharmonic Jaynes-Cummings ladder for a two-level qubit. The observation of photon antibunching dips ($g^{(2)}(0) < 1$) at the polariton resonances as well as the bunching peaks ($g^{(2)}(0) > 1$) at the energies of the two second-manifold states has been extensively explored in atomic \cite{birnbaum2005photon} and solid-state systems \cite{Photon_Blockade_AF}. 

When U becomes comparable to $g$, there appears another energy, separate from the polaritons, that produces antibunching. As explained by Bamba \textit{et al.} \cite{bamba2011origin}, this antibunching dip is a result of destructive quantum interference between the first and the second manifolds, and its energy is given by      
\begin{equation}
2\omega'^3 + 2 U \omega'^2 + g^2 U = 0
 \end{equation} 
where $\omega' = \omega - i\frac{\Gamma}{2}$.

Figure 2 shows a plot of $g^{(2)}(0)$ versus the pump laser frequency detuned from the cavity resonance at multiple values of U. In addition to the first and the second-manifold eigenenergies plotted in Fig. 1c, the interference-induced antibunching is clearly observed in Fig. 2a (the color represents the base-10 logarithm of $g^{(2)}(0)$). As U increases, the interference dip passes through the upper polariton dip at $\omega_{pump} = g$. Figure. 2b shows the cross-sections of Fig. 2a for U/$g$ = 0.3, 0.67, and 1.5. For U/$g$ = 0.67 (shown in green), the interference dip coincides with the upper polariton dip, yielding an extremely strong antibunching ($g^{(2)}(0) \sim 10^{-7}$) .

\begin{figure}
\includegraphics[width=85mm]{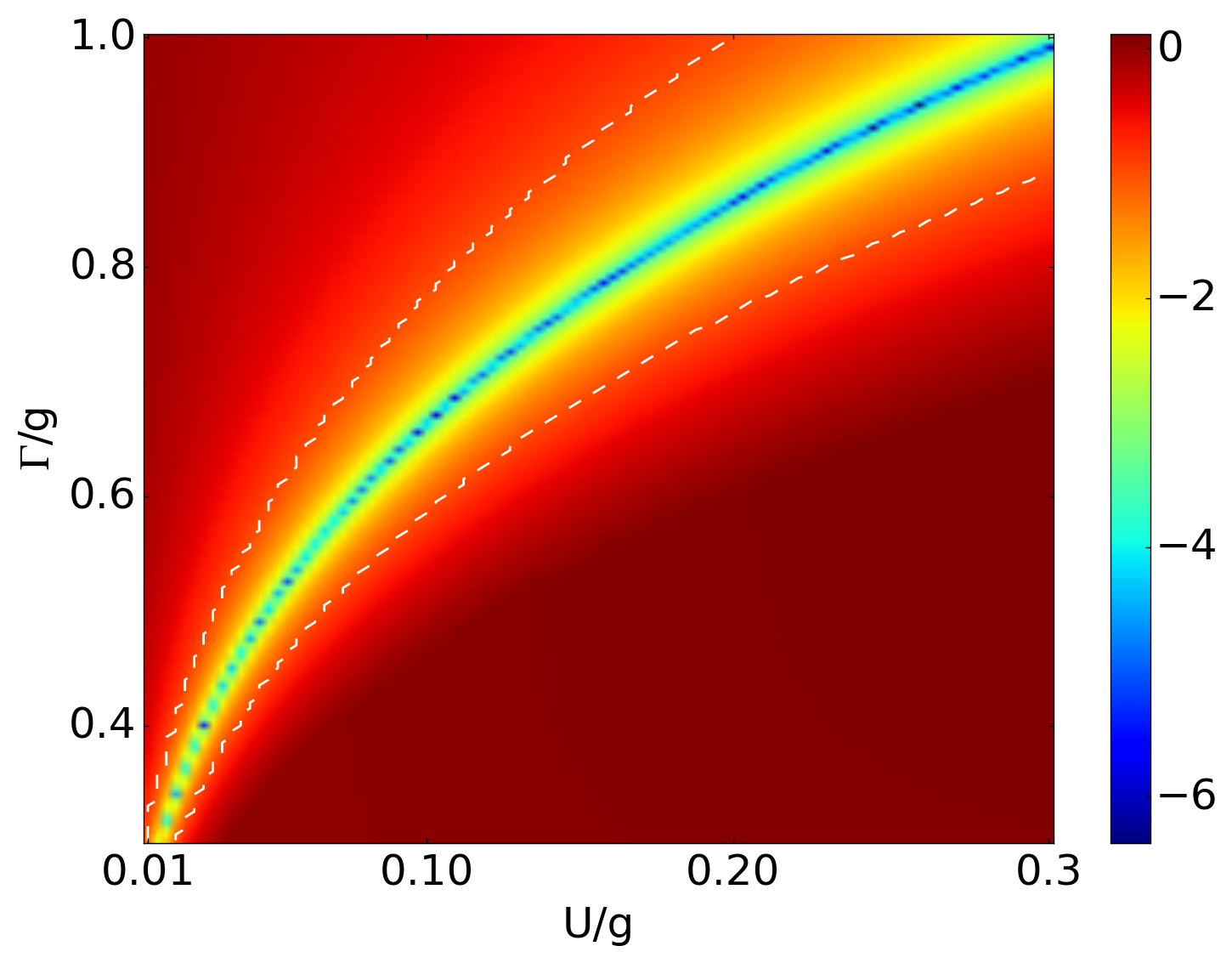}
\caption{\label{Figure:g2_vs_U_vs_Gamma_colorplot_kappa} (Color online) \textbf{Minimum $g^{(2)}(0)$ for different $\Gamma$ and U}. A 2D plot of the minimum value of the $g^{(2)}(0)$ that appears at negative $\omega_{pump}$ (see Fig. 3) versus U (x-axis) and $\Gamma$ (y-axis). The color represents the base-10 logarithm of $g^{(2)}(0)$. For a given value of $\Gamma$, strong antibunching is observed for a range of U. As $\Gamma$ increases, the optimal value of U as well as its width increase. White dashed lines mark where $g^{(2)}(0)$ = 0.1. For this simulation, $\kappa$ is set at 0.5g. The dotted appearance for strong antibunching is a numerical artifact.}
\end{figure}

\begin{figure}
\includegraphics[width=90mm]{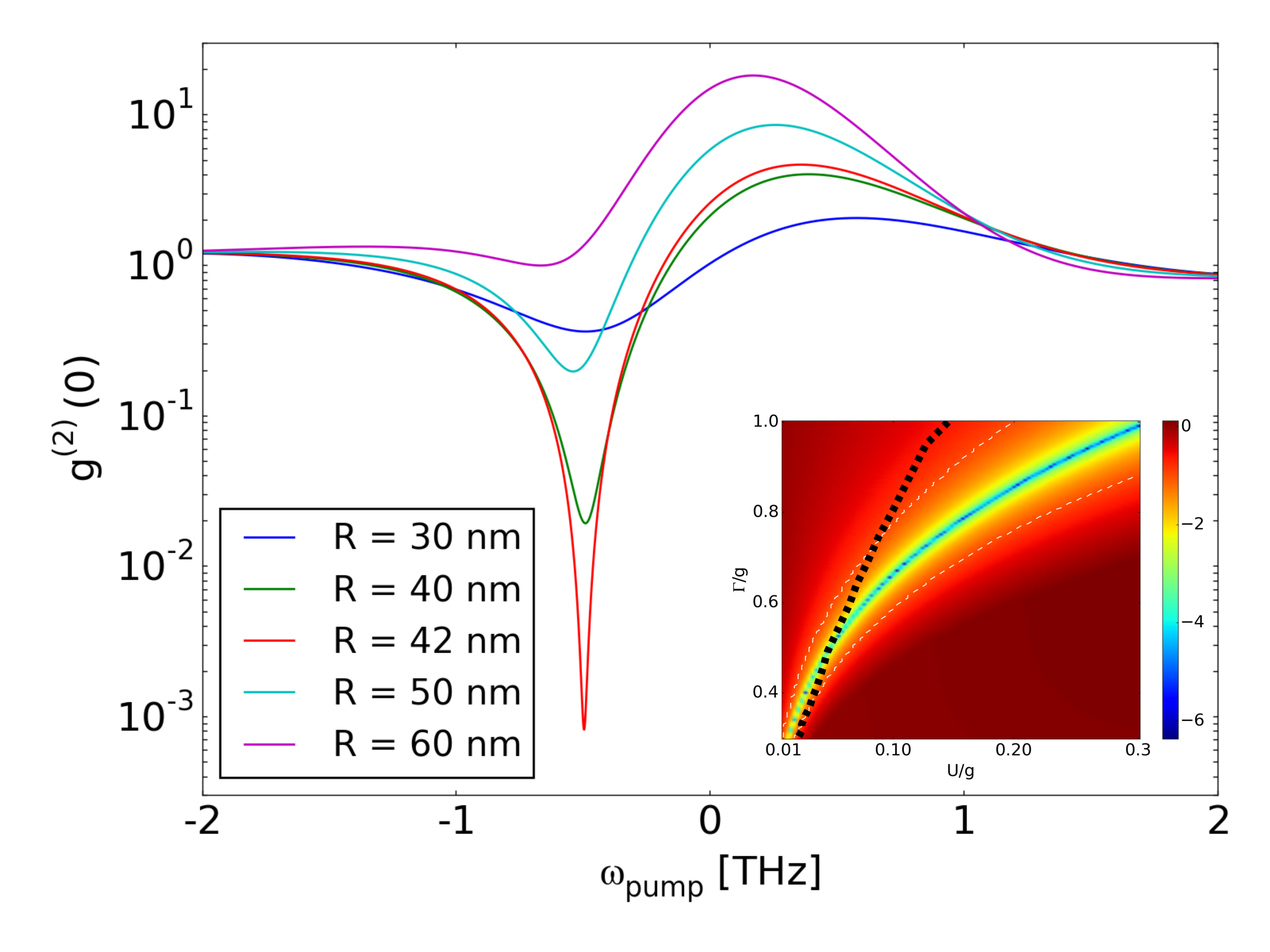}
\caption{\label{Figure:plot_g2_vs_detuning_S} (Color online) \textbf{$g^{(2)}(0)$ vs pump laser frequency for different S}. A plot of $g^{(2)}(0)$ versus pump laser frequency detuning for different monolayer area, with radius $R$ ranging from 30 nm to 60 nm. The strong antibunching appears for $R$ = 42 nm. (inset) The effect of $R$ on the other parameters $g$, U, and $\Gamma$, and consequently $g^{(2)}(0)$, can be seen by plotting the parameters (black dotted line) on top of Fig. 4. As $R$ is changed, the set of parameters cuts across the region of strong antibunching, making the system dynamics tunable.}
\end{figure}

Having explored $\Gamma \ll g \sim U$, we increase the dissipation in our system until it becomes comparable to $g$, which is more representative of typical solid-state environments. In Fig. 3, we explore three separate values of $\Gamma$/$g$: 0.1, 0.5, and 1.0. For each one, we plot $g^{(2)}(0)$ versus the pump laser detuning for a range of U values. As $\Gamma$ increases, previously sharp features become rounded, and what used to be a strong antibunching dip at $\omega_{pump} = g$ becomes gradually shallower (Fig. 3a).

For large $\Gamma$, on the other hand, an additional antibunching dip appears. As seen in Fig. 3b and c, this dip only appears for U $< \Gamma$, and the value of U at which it appears depends on how close $\Gamma$/g is to unity. The origin of this antibunching is once again the destructive quantum interference \cite{bamba2011origin}, which has been extensively investigated by Liew \textit{et al.} in the context of ``polariton boxes'' \cite{liew2010single}. For a given $\Gamma$, Eq. III.1 gives the optimum U and $\omega$ that produce the smallest $g^{(2)}(0)$.

Figure 4 displays a two-dimensional color plot of minimum $g^{(2)}(0)$ as a function of $\Gamma$ and U. Here we set $\kappa = 0.5g$. The color represents the base-10 logarithm of $g^{(2)}(0)$, ranging from red ($g^{(2)}(0) \approx 1$) to blue ($g^{(2)}(0) \approx 10^{-6}$). We have indicated on the plot with white dotted lines where $g^{(2)}(0) = 0.1$, showing that the domain of U that produces strong antibunching increases with $\Gamma$. 


\section{\label{sec:Para1} Proposed experimental design}

To observe the strong, interference-induced antibunching, we propose to pattern a 2D-material monolayer into a circular island with radius $R$ and place it on a thin photonic crystal cavity (see Fig. 1a). We assume that the area of the patterned monolayer is much smaller than that of the cavity mode, i.e., $R\ll R_{mode}$. We also assume that the monolayer is free of any defect such that the excitons are delocalized over the entire monolayer area. Hence, the spatial extent of the exciton wavefunction is equal to the physical size of the monolayer.

Both the exciton-photon coupling $g$ and the nonlinearity U depend on the size of the monolayer. The former is given by \cite{wang2017quantum}
\begin{equation}
\hbar g = \frac{d_{cv} |\phi(0)| \sqrt{\hbar \omega_c}}{\sqrt{2\epsilon_0 L_c}} \sqrt{\frac{\pi R^2}{\pi R_{mode}^2}}
\end{equation}
where $d_{cv}$ is the interband dipole matrix element, $|\phi(0)| = \sqrt{2/(\pi a_B)^2}$ is the amplitude of the exciton wavefunction ($a_B$ is the exciton Bohr radius), $\omega_c$ is the cavity resonance frequency, $\epsilon_0$ is the permittivity of free space, and $L_c$ is the effective length of the cavity mode. The nonlinear interaction strength is given by $U = 6 E_b a_B^2/(\pi R^2)$, where $E_b$ is the exciton binding energy \cite{tassone1999exciton}.

Thus, $g \sim R$ and U $\sim 1/R^2$, allowing us to tune the system dynamics by patterning the monolayer into different areas via, for instance, electron beam lithography. For a W$\text{Se}_2$ monolayer with $R=5$ nm coupled to a SiN nanobeam cavity with $R_{mode} = 1$ $\mu$m, $g \approx 2\pi \times 700$ GHz and U $\approx 2\pi \times 30$ GHz \cite{wang2017quantum}.

\begin{figure}
\includegraphics[width=85mm]{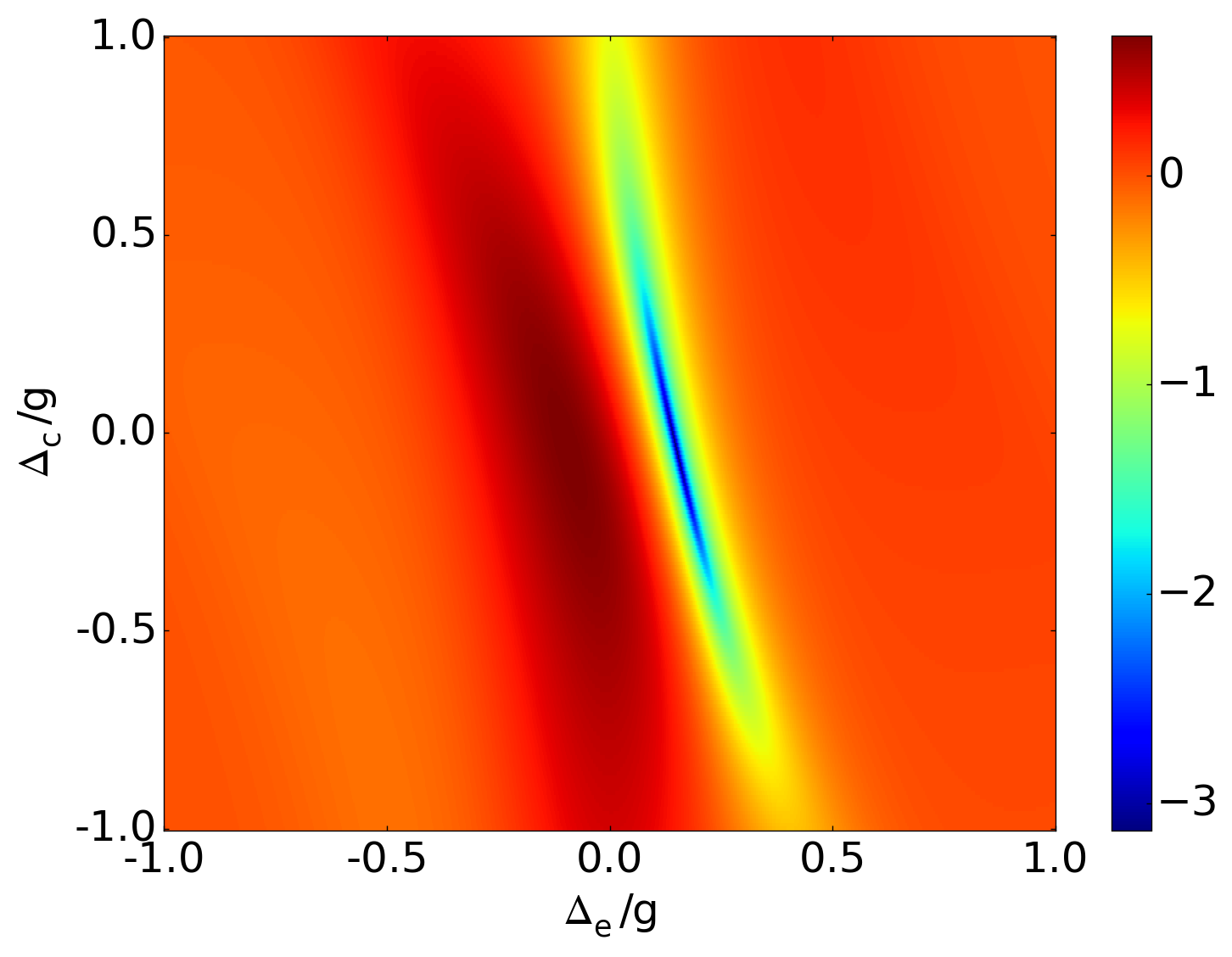}
\caption{\label{Figure:two_level} (Color online) \textbf{$g^{(2)}(0)$ vs $\Delta_e$ vs $\Delta_c$}. A 2D plot of $g^{(2)}(0)$ versus exciton detuning (x-axis) and cavity detuning (y-axis) for monolayer radius $R$ = 42 nm. The color represents the base-10 logarithm of $g^{(2)}(0)$. Clearly, the variance is much greater for the exciton detuning compared to that for the cavity detuning.}
\end{figure}

While the cavity loss for a typical nanobeam is fixed ($\kappa = 2\pi \times 150$ GHz) \cite{fryett2017encapsulated}, the exact dependence of $\Gamma$ on $R$ is unknown and remains an open problem. It has been reported that patterned monolayers on the order of tens of nanometers in radii can suffer from linewidth broadening due to the presence of edge states. Since the length of the edge scales linearly with R and the loss has been seen to increase for smaller monolayers, for our simulations, we have chosen to fix $\Gamma = 2\pi \times 300$ GHz at $R = 50$ nm, an experimentally measured value, and vary it as $1/R$ \cite{fryett2017encapsulated}.  

Figure 5 shows the effect of changing $R$ on $g^{(2)}(0)$. As $R$ increases from 30 nm to 60 nm, an antibunching dip appears, becomes sharper, and then recedes. The strongest antibunching occurs at $R = 42$ nm. The inset shows how the appearance of the dip compares to the general antibunching behavior in Fig. 4. The black dotted line, representing changing $R$, cuts across the region of strong antibunching, exhibiting the system's tunability.


Finally, we explore the robustness of the antibunching dip for unequal cavity and exciton detunings, i.e., $\omega_c \ne \omega_e$. Figure 6 shows a plot of $g^{(2)}(0)$ as a function of $\omega_c$ and $\omega_e$ for the optimal paramters ($R=42$ nm, $g = 2\pi \times 560$ GHz, $\Gamma = 2\pi \times 360$ GHz, $\kappa = 2\pi \times 150$ GHz, U = $2\pi \times 40$ GHz), where the color represents the base-10 logarithm of $g^{(2)}(0)$. While the antibunching behavior is observed only for a narrow range of the exciton detuning (x-axis), it survives for a much larger range of the cavity detuning (y-axis), giving us substantial leeway in the fabrication precision of the nanobeam cavity.

\section{\label{sec:Outlook} Conclusion}

We have explored the second-order coherence of a 2D-material monolayer embedded in a photonic crystal cavity and identified a range of system parameters that yield strong photon antibunching. We have shown that by patterning the monolayer into different sizes, we can tune the system dynamics, driving it from a weak to a strong photon antibunching regime. The successful implementation of the experimental design will open the door to a new regime of quantum interference-based quantum simulations on a scalable, on-chip platform.


\section{\label{sec:Acknowledgements}Acknowledgements}
This work was supported by the National Science Foundation under grants NSF-EFRI-1433496 and NSF-1708579 and the Air Force Office of Scientific Research-Young Investigator Program under grant FA9550-15-1-0150.



\bibliography{TheBib}{}

\begin{thebibliography}{34}%
\makeatletter
\providecommand \@ifxundefined [1]{%
 \@ifx{#1\undefined}
}%
\providecommand \@ifnum [1]{%
 \ifnum #1\expandafter \@firstoftwo
 \else \expandafter \@secondoftwo
 \fi
}%
\providecommand \@ifx [1]{%
 \ifx #1\expandafter \@firstoftwo
 \else \expandafter \@secondoftwo
 \fi
}%
\providecommand \natexlab [1]{#1}%
\providecommand \enquote  [1]{``#1''}%
\providecommand \bibnamefont  [1]{#1}%
\providecommand \bibfnamefont [1]{#1}%
\providecommand \citenamefont [1]{#1}%
\providecommand \href@noop [0]{\@secondoftwo}%
\providecommand \href [0]{\begingroup \@sanitize@url \@href}%
\providecommand \@href[1]{\@@startlink{#1}\@@href}%
\providecommand \@@href[1]{\endgroup#1\@@endlink}%
\providecommand \@sanitize@url [0]{\catcode `\\12\catcode `\$12\catcode
  `\&12\catcode `\#12\catcode `\^12\catcode `\_12\catcode `\%12\relax}%
\providecommand \@@startlink[1]{}%
\providecommand \@@endlink[0]{}%
\providecommand \url  [0]{\begingroup\@sanitize@url \@url }%
\providecommand \@url [1]{\endgroup\@href {#1}{\urlprefix }}%
\providecommand \urlprefix  [0]{URL }%
\providecommand \Eprint [0]{\href }%
\providecommand \doibase [0]{http://dx.doi.org/}%
\providecommand \selectlanguage [0]{\@gobble}%
\providecommand \bibinfo  [0]{\@secondoftwo}%
\providecommand \bibfield  [0]{\@secondoftwo}%
\providecommand \translation [1]{[#1]}%
\providecommand \BibitemOpen [0]{}%
\providecommand \bibitemStop [0]{}%
\providecommand \bibitemNoStop [0]{.\EOS\space}%
\providecommand \EOS [0]{\spacefactor3000\relax}%
\providecommand \BibitemShut  [1]{\csname bibitem#1\endcsname}%
\let\auto@bib@innerbib\@empty
\bibitem [{\citenamefont {Kiraz}\ \emph {et~al.}(2004)\citenamefont {Kiraz},
  \citenamefont {Atat\"ure},\ and\ \citenamefont {Imamo\ifmmode~\breve{g}\else
  \u{g}\fi{}lu}}]{kiraz2004quantum}%
  \BibitemOpen
  \bibfield  {author} {\bibinfo {author} {\bibfnamefont {A.}~\bibnamefont
  {Kiraz}}, \bibinfo {author} {\bibfnamefont {M.}~\bibnamefont {Atat\"ure}}, \
  and\ \bibinfo {author} {\bibfnamefont {A.}~\bibnamefont
  {Imamo\ifmmode~\breve{g}\else \u{g}\fi{}lu}},\ }\href {\doibase
  10.1103/PhysRevA.69.032305} {\bibfield  {journal} {\bibinfo  {journal} {Phys.
  Rev. A}\ }\textbf {\bibinfo {volume} {69}},\ \bibinfo {pages} {032305}
  (\bibinfo {year} {2004})}\BibitemShut {NoStop}%
\bibitem [{\citenamefont {Kuhn}\ \emph {et~al.}(2002)\citenamefont {Kuhn},
  \citenamefont {Hennrich},\ and\ \citenamefont
  {Rempe}}]{kuhn2002deterministic}%
  \BibitemOpen
  \bibfield  {author} {\bibinfo {author} {\bibfnamefont {A.}~\bibnamefont
  {Kuhn}}, \bibinfo {author} {\bibfnamefont {M.}~\bibnamefont {Hennrich}}, \
  and\ \bibinfo {author} {\bibfnamefont {G.}~\bibnamefont {Rempe}},\ }\href
  {\doibase 10.1103/PhysRevLett.89.067901} {\bibfield  {journal} {\bibinfo
  {journal} {Phys. Rev. Lett.}\ }\textbf {\bibinfo {volume} {89}},\ \bibinfo
  {pages} {067901} (\bibinfo {year} {2002})}\BibitemShut {NoStop}%
\bibitem [{\citenamefont {Milburn}(1989)}]{milburn1989quantum}%
  \BibitemOpen
  \bibfield  {author} {\bibinfo {author} {\bibfnamefont {G.~J.}\ \bibnamefont
  {Milburn}},\ }\href {\doibase 10.1103/PhysRevLett.62.2124} {\bibfield
  {journal} {\bibinfo  {journal} {Phys. Rev. Lett.}\ }\textbf {\bibinfo
  {volume} {62}},\ \bibinfo {pages} {2124} (\bibinfo {year}
  {1989})}\BibitemShut {NoStop}%
\bibitem [{\citenamefont {Bakr}\ \emph {et~al.}(2009)\citenamefont {Bakr},
  \citenamefont {Gillen}, \citenamefont {Peng}, \citenamefont {F{\"o}lling},\
  and\ \citenamefont {Greiner}}]{bakr2009quantum}%
  \BibitemOpen
  \bibfield  {author} {\bibinfo {author} {\bibfnamefont {W.~S.}\ \bibnamefont
  {Bakr}}, \bibinfo {author} {\bibfnamefont {J.~I.}\ \bibnamefont {Gillen}},
  \bibinfo {author} {\bibfnamefont {A.}~\bibnamefont {Peng}}, \bibinfo {author}
  {\bibfnamefont {S.}~\bibnamefont {F{\"o}lling}}, \ and\ \bibinfo {author}
  {\bibfnamefont {M.}~\bibnamefont {Greiner}},\ }\href {\doibase
  10.1038/nature08482} {\bibfield  {journal} {\bibinfo  {journal} {Nature
  (London)}\ }\textbf {\bibinfo {volume} {462}},\ \bibinfo {pages} {74}
  (\bibinfo {year} {2009})}\BibitemShut {NoStop}%
\bibitem [{\citenamefont {Devoret}\ and\ \citenamefont
  {Schoelkopf}(2013)}]{devoret2013superconducting}%
  \BibitemOpen
  \bibfield  {author} {\bibinfo {author} {\bibfnamefont {M.~H.}\ \bibnamefont
  {Devoret}}\ and\ \bibinfo {author} {\bibfnamefont {R.~J.}\ \bibnamefont
  {Schoelkopf}},\ }\href {\doibase 10.1126/science.1231930} {\bibfield
  {journal} {\bibinfo  {journal} {Science}\ }\textbf {\bibinfo {volume}
  {339}},\ \bibinfo {pages} {1169} (\bibinfo {year} {2013})}\BibitemShut
  {NoStop}%
\bibitem [{\citenamefont {Majumdar}\ \emph {et~al.}(2012)\citenamefont
  {Majumdar}, \citenamefont {Bajcsy},\ and\ \citenamefont {Vu\ifmmode
  \check{c}\else \v{c}\fi{}kovi\ifmmode~\acute{c}\else
  \'{c}\fi{}}}]{Photon_Blockade_AM}%
  \BibitemOpen
  \bibfield  {author} {\bibinfo {author} {\bibfnamefont {A.}~\bibnamefont
  {Majumdar}}, \bibinfo {author} {\bibfnamefont {M.}~\bibnamefont {Bajcsy}}, \
  and\ \bibinfo {author} {\bibfnamefont {J.}~\bibnamefont {Vu\ifmmode
  \check{c}\else \v{c}\fi{}kovi\ifmmode~\acute{c}\else \'{c}\fi{}}},\ }\href
  {\doibase 10.1103/PhysRevA.85.041801} {\bibfield  {journal} {\bibinfo
  {journal} {Phys. Rev. A}\ }\textbf {\bibinfo {volume} {85}},\ \bibinfo
  {pages} {041801} (\bibinfo {year} {2012})}\BibitemShut {NoStop}%
\bibitem [{\citenamefont {Faraon}\ \emph {et~al.}(2008)\citenamefont {Faraon},
  \citenamefont {Fushman}, \citenamefont {Englund}, \citenamefont {Stoltz},
  \citenamefont {Petroff},\ and\ \citenamefont
  {Vu{\v{c}}kovi{\'{c}}}}]{Photon_Blockade_AF}%
  \BibitemOpen
  \bibfield  {author} {\bibinfo {author} {\bibfnamefont {A.}~\bibnamefont
  {Faraon}}, \bibinfo {author} {\bibfnamefont {I.}~\bibnamefont {Fushman}},
  \bibinfo {author} {\bibfnamefont {D.}~\bibnamefont {Englund}}, \bibinfo
  {author} {\bibfnamefont {N.}~\bibnamefont {Stoltz}}, \bibinfo {author}
  {\bibfnamefont {P.}~\bibnamefont {Petroff}}, \ and\ \bibinfo {author}
  {\bibfnamefont {J.}~\bibnamefont {Vu{\v{c}}kovi{\'{c}}}},\ }\href {\doibase
  10.1038/nphys1078} {\bibfield  {journal} {\bibinfo  {journal} {Nat. Phys.}\
  }\textbf {\bibinfo {volume} {4}},\ \bibinfo {pages} {859} (\bibinfo {year}
  {2008})}\BibitemShut {NoStop}%
\bibitem [{\citenamefont {Reinhard}\ \emph {et~al.}(2012)\citenamefont
  {Reinhard}, \citenamefont {Volz}, \citenamefont {Winger}, \citenamefont
  {Badolato}, \citenamefont {Hennessy}, \citenamefont {Hu},\ and\ \citenamefont
  {Imamoglu}}]{Photon_Blockade_Atac}%
  \BibitemOpen
  \bibfield  {author} {\bibinfo {author} {\bibfnamefont {A.}~\bibnamefont
  {Reinhard}}, \bibinfo {author} {\bibfnamefont {T.}~\bibnamefont {Volz}},
  \bibinfo {author} {\bibfnamefont {M.}~\bibnamefont {Winger}}, \bibinfo
  {author} {\bibfnamefont {A.}~\bibnamefont {Badolato}}, \bibinfo {author}
  {\bibfnamefont {K.~J.}\ \bibnamefont {Hennessy}}, \bibinfo {author}
  {\bibfnamefont {E.~L.}\ \bibnamefont {Hu}}, \ and\ \bibinfo {author}
  {\bibfnamefont {A.}~\bibnamefont {Imamoglu}},\ }\href {\doibase
  10.1038/nphoton2011321} {\bibfield  {journal} {\bibinfo  {journal} {Nat.
  Photonics}\ }\textbf {\bibinfo {volume} {6}},\ \bibinfo {pages} {93}
  (\bibinfo {year} {2012})}\BibitemShut {NoStop}%
\bibitem [{\citenamefont {Negoita}\ \emph {et~al.}(1999)\citenamefont
  {Negoita}, \citenamefont {Snoke},\ and\ \citenamefont {Eberl}}]{QW_strain}%
  \BibitemOpen
  \bibfield  {author} {\bibinfo {author} {\bibfnamefont {V.}~\bibnamefont
  {Negoita}}, \bibinfo {author} {\bibfnamefont {D.~W.}\ \bibnamefont {Snoke}},
  \ and\ \bibinfo {author} {\bibfnamefont {K.}~\bibnamefont {Eberl}},\ }\href
  {\doibase 10.1063/1.124915} {\bibfield  {journal} {\bibinfo  {journal}
  {Applied Physics Letters}\ }\textbf {\bibinfo {volume} {75}},\ \bibinfo
  {pages} {2059} (\bibinfo {year} {1999})}\BibitemShut {NoStop}%
\bibitem [{\citenamefont {Hartmann}(2016)}]{Quantum_simulation_CCA_Hartman}%
  \BibitemOpen
  \bibfield  {author} {\bibinfo {author} {\bibfnamefont {M.~J.}\ \bibnamefont
  {Hartmann}},\ }\href {http://stacks.iop.org/2040-8986/18/i=10/a=104005}
  {\bibfield  {journal} {\bibinfo  {journal} {Journal of Optics}\ }\textbf
  {\bibinfo {volume} {18}},\ \bibinfo {pages} {104005} (\bibinfo {year}
  {2016})}\BibitemShut {NoStop}%
\bibitem [{\citenamefont {Kim}\ and\ \citenamefont
  {Yamamoto}(2017)}]{Na_Young_Kim_EP_Simulator}%
  \BibitemOpen
  \bibfield  {author} {\bibinfo {author} {\bibfnamefont {N.~Y.}\ \bibnamefont
  {Kim}}\ and\ \bibinfo {author} {\bibfnamefont {Y.}~\bibnamefont {Yamamoto}},\
  }\enquote {\bibinfo {title} {Exciton-polariton quantum simulators},}\ in\
  \href {\doibase 10.1007/978-3-319-52025-4_5} {\emph {\bibinfo {booktitle}
  {Quantum Simulations with Photons and Polaritons: Merging Quantum Optics with
  Condensed Matter Physics}}},\ \bibinfo {editor} {edited by\ \bibinfo {editor}
  {\bibfnamefont {D.~G.}\ \bibnamefont {Angelakis}}}\ (\bibinfo  {publisher}
  {Springer International Publishing},\ \bibinfo {address} {Cham},\ \bibinfo
  {year} {2017})\ pp.\ \bibinfo {pages} {91--121}\BibitemShut {NoStop}%
\bibitem [{\citenamefont {Carusotto}\ and\ \citenamefont
  {Ciuti}(2013)}]{Quantum_fluids_light}%
  \BibitemOpen
  \bibfield  {author} {\bibinfo {author} {\bibfnamefont {I.}~\bibnamefont
  {Carusotto}}\ and\ \bibinfo {author} {\bibfnamefont {C.}~\bibnamefont
  {Ciuti}},\ }\href {\doibase 10.1103/RevModPhys.85.299} {\bibfield  {journal}
  {\bibinfo  {journal} {Rev. Mod. Phys.}\ }\textbf {\bibinfo {volume} {85}},\
  \bibinfo {pages} {299} (\bibinfo {year} {2013})}\BibitemShut {NoStop}%
\bibitem [{\citenamefont {Carusotto}\ \emph {et~al.}(2009)\citenamefont
  {Carusotto}, \citenamefont {Gerace}, \citenamefont {Tureci}, \citenamefont
  {De~Liberato}, \citenamefont {Ciuti},\ and\ \citenamefont
  {Imamo\ifmmode~\check{g}\else \v{g}\fi{}lu}}]{Fermionized_Photons}%
  \BibitemOpen
  \bibfield  {author} {\bibinfo {author} {\bibfnamefont {I.}~\bibnamefont
  {Carusotto}}, \bibinfo {author} {\bibfnamefont {D.}~\bibnamefont {Gerace}},
  \bibinfo {author} {\bibfnamefont {H.~E.}\ \bibnamefont {Tureci}}, \bibinfo
  {author} {\bibfnamefont {S.}~\bibnamefont {De~Liberato}}, \bibinfo {author}
  {\bibfnamefont {C.}~\bibnamefont {Ciuti}}, \ and\ \bibinfo {author}
  {\bibfnamefont {A.}~\bibnamefont {Imamo\ifmmode~\check{g}\else
  \v{g}\fi{}lu}},\ }\href {\doibase 10.1103/PhysRevLett.103.033601} {\bibfield
  {journal} {\bibinfo  {journal} {Phys. Rev. Lett.}\ }\textbf {\bibinfo
  {volume} {103}},\ \bibinfo {pages} {033601} (\bibinfo {year}
  {2009})}\BibitemShut {NoStop}%
\bibitem [{\citenamefont {Deng}\ \emph {et~al.}(2010)\citenamefont {Deng},
  \citenamefont {Haug},\ and\ \citenamefont {Yamamoto}}]{Hui_deng_EP_BEC}%
  \BibitemOpen
  \bibfield  {author} {\bibinfo {author} {\bibfnamefont {H.}~\bibnamefont
  {Deng}}, \bibinfo {author} {\bibfnamefont {H.}~\bibnamefont {Haug}}, \ and\
  \bibinfo {author} {\bibfnamefont {Y.}~\bibnamefont {Yamamoto}},\ }\href
  {\doibase 10.1103/RevModPhys.82.1489} {\bibfield  {journal} {\bibinfo
  {journal} {Rev. Mod. Phys.}\ }\textbf {\bibinfo {volume} {82}},\ \bibinfo
  {pages} {1489} (\bibinfo {year} {2010})}\BibitemShut {NoStop}%
\bibitem [{\citenamefont {Sanvitto}\ and\ \citenamefont
  {K{\'{e}}na-Cohen}(2016)}]{Polariton_devices}%
  \BibitemOpen
  \bibfield  {author} {\bibinfo {author} {\bibfnamefont {D.}~\bibnamefont
  {Sanvitto}}\ and\ \bibinfo {author} {\bibfnamefont {S.}~\bibnamefont
  {K{\'{e}}na-Cohen}},\ }\href {\doibase 10.1038/nmat4668} {\bibfield
  {journal} {\bibinfo  {journal} {Nat. Mater.}\ }\textbf {\bibinfo {volume}
  {15}},\ \bibinfo {pages} {1061} (\bibinfo {year} {2016})}\BibitemShut
  {NoStop}%
\bibitem [{\citenamefont {Besga}\ \emph {et~al.}(2015)\citenamefont {Besga},
  \citenamefont {Vaneph}, \citenamefont {Reichel}, \citenamefont {Est\`eve},
  \citenamefont {Reinhard}, \citenamefont {Miguel-S\'anchez}, \citenamefont
  {Imamo\ifmmode~\breve{g}\else \u{g}\fi{}lu},\ and\ \citenamefont
  {Volz}}]{besga2015polariton}%
  \BibitemOpen
  \bibfield  {author} {\bibinfo {author} {\bibfnamefont {B.}~\bibnamefont
  {Besga}}, \bibinfo {author} {\bibfnamefont {C.}~\bibnamefont {Vaneph}},
  \bibinfo {author} {\bibfnamefont {J.}~\bibnamefont {Reichel}}, \bibinfo
  {author} {\bibfnamefont {J.}~\bibnamefont {Est\`eve}}, \bibinfo {author}
  {\bibfnamefont {A.}~\bibnamefont {Reinhard}}, \bibinfo {author}
  {\bibfnamefont {J.}~\bibnamefont {Miguel-S\'anchez}}, \bibinfo {author}
  {\bibfnamefont {A.}~\bibnamefont {Imamo\ifmmode~\breve{g}\else
  \u{g}\fi{}lu}}, \ and\ \bibinfo {author} {\bibfnamefont {T.}~\bibnamefont
  {Volz}},\ }\href {\doibase 10.1103/PhysRevApplied.3.014008} {\bibfield
  {journal} {\bibinfo  {journal} {Phys. Rev. Applied}\ }\textbf {\bibinfo
  {volume} {3}},\ \bibinfo {pages} {014008} (\bibinfo {year}
  {2015})}\BibitemShut {NoStop}%
\bibitem [{\citenamefont {Mu{\~n}oz-Matutano}\ \emph
  {et~al.}(2017)\citenamefont {Mu{\~n}oz-Matutano}, \citenamefont {Wood},
  \citenamefont {Johnson}, \citenamefont {Asensio}, \citenamefont {Baragiola},
  \citenamefont {Reinhard}, \citenamefont {Lemaitre}, \citenamefont {Bloch},
  \citenamefont {Amo}, \citenamefont {Besga} \emph
  {et~al.}}]{munoz2017quantum}%
  \BibitemOpen
  \bibfield  {author} {\bibinfo {author} {\bibfnamefont {G.}~\bibnamefont
  {Mu{\~n}oz-Matutano}}, \bibinfo {author} {\bibfnamefont {A.}~\bibnamefont
  {Wood}}, \bibinfo {author} {\bibfnamefont {M.}~\bibnamefont {Johnson}},
  \bibinfo {author} {\bibfnamefont {X.~V.}\ \bibnamefont {Asensio}}, \bibinfo
  {author} {\bibfnamefont {B.}~\bibnamefont {Baragiola}}, \bibinfo {author}
  {\bibfnamefont {A.}~\bibnamefont {Reinhard}}, \bibinfo {author}
  {\bibfnamefont {A.}~\bibnamefont {Lemaitre}}, \bibinfo {author}
  {\bibfnamefont {J.}~\bibnamefont {Bloch}}, \bibinfo {author} {\bibfnamefont
  {A.}~\bibnamefont {Amo}}, \bibinfo {author} {\bibfnamefont {B.}~\bibnamefont
  {Besga}},  \emph {et~al.},\ }\href@noop {} {\bibfield  {journal} {\bibinfo
  {journal} {arXiv preprint arXiv:1712.05551}\ } (\bibinfo {year}
  {2017})}\BibitemShut {NoStop}%
\bibitem [{\citenamefont {Verma}\ and\ \citenamefont
  {Coleman}(2008)}]{verma2008high}%
  \BibitemOpen
  \bibfield  {author} {\bibinfo {author} {\bibfnamefont {V.}~\bibnamefont
  {Verma}}\ and\ \bibinfo {author} {\bibfnamefont {J.}~\bibnamefont
  {Coleman}},\ }\href {\doibase 10.1063/1.2981207} {\bibfield  {journal}
  {\bibinfo  {journal} {Applied Physics Letters}\ }\textbf {\bibinfo {volume}
  {93}},\ \bibinfo {pages} {111117} (\bibinfo {year} {2008})}\BibitemShut
  {NoStop}%
\bibitem [{\citenamefont {Lee}\ \emph {et~al.}(2011)\citenamefont {Lee},
  \citenamefont {Zhang}, \citenamefont {Deng},\ and\ \citenamefont
  {Ku}}]{lee2011room}%
  \BibitemOpen
  \bibfield  {author} {\bibinfo {author} {\bibfnamefont {L.}~\bibnamefont
  {Lee}}, \bibinfo {author} {\bibfnamefont {L.}~\bibnamefont {Zhang}}, \bibinfo
  {author} {\bibfnamefont {H.}~\bibnamefont {Deng}}, \ and\ \bibinfo {author}
  {\bibfnamefont {P.-C.}\ \bibnamefont {Ku}},\ }\href {\doibase
  10.1063/1.3672441} {\bibfield  {journal} {\bibinfo  {journal} {Applied
  Physics Letters}\ }\textbf {\bibinfo {volume} {99}},\ \bibinfo {pages}
  {263105} (\bibinfo {year} {2011})}\BibitemShut {NoStop}%
\bibitem [{\citenamefont {Fryett}\ \emph
  {et~al.}(2017{\natexlab{a}})\citenamefont {Fryett}, \citenamefont {Zhan},\
  and\ \citenamefont {Majumdar}}]{fryett2017cavity}%
  \BibitemOpen
  \bibfield  {author} {\bibinfo {author} {\bibfnamefont {T.}~\bibnamefont
  {Fryett}}, \bibinfo {author} {\bibfnamefont {A.}~\bibnamefont {Zhan}}, \ and\
  \bibinfo {author} {\bibfnamefont {A.}~\bibnamefont {Majumdar}},\ }\href@noop
  {} {\bibfield  {journal} {\bibinfo  {journal} {Nanophotonics}\ } (\bibinfo
  {year} {2017}{\natexlab{a}})}\BibitemShut {NoStop}%
\bibitem [{\citenamefont {Wang}\ \emph {et~al.}(2012)\citenamefont {Wang},
  \citenamefont {Kalantar-Zadeh}, \citenamefont {Kis}, \citenamefont
  {Coleman},\ and\ \citenamefont {Strano}}]{TMDC_optoelectronics}%
  \BibitemOpen
  \bibfield  {author} {\bibinfo {author} {\bibfnamefont {Q.~H.}\ \bibnamefont
  {Wang}}, \bibinfo {author} {\bibfnamefont {K.}~\bibnamefont
  {Kalantar-Zadeh}}, \bibinfo {author} {\bibfnamefont {A.}~\bibnamefont {Kis}},
  \bibinfo {author} {\bibfnamefont {J.~N.}\ \bibnamefont {Coleman}}, \ and\
  \bibinfo {author} {\bibfnamefont {M.~S.}\ \bibnamefont {Strano}},\ }\href
  {\doibase 10.1038/nnano.2012.193} {\bibfield  {journal} {\bibinfo  {journal}
  {Nat. Nanotechnol.}\ }\textbf {\bibinfo {volume} {7}},\ \bibinfo {pages}
  {699} (\bibinfo {year} {2012})}\BibitemShut {NoStop}%
\bibitem [{\citenamefont {Wu}\ \emph {et~al.}(2015)\citenamefont {Wu},
  \citenamefont {Buckley}, \citenamefont {Schaibley}, \citenamefont {Feng},
  \citenamefont {Yan}, \citenamefont {Mandrus}, \citenamefont {Hatami},
  \citenamefont {Yao}, \citenamefont {Vuckovic}, \citenamefont {Majumdar},\
  and\ \citenamefont {Xu}}]{Laser_AM_Xu}%
  \BibitemOpen
  \bibfield  {author} {\bibinfo {author} {\bibfnamefont {S.}~\bibnamefont
  {Wu}}, \bibinfo {author} {\bibfnamefont {S.}~\bibnamefont {Buckley}},
  \bibinfo {author} {\bibfnamefont {J.~R.}\ \bibnamefont {Schaibley}}, \bibinfo
  {author} {\bibfnamefont {L.}~\bibnamefont {Feng}}, \bibinfo {author}
  {\bibfnamefont {J.}~\bibnamefont {Yan}}, \bibinfo {author} {\bibfnamefont
  {D.~G.}\ \bibnamefont {Mandrus}}, \bibinfo {author} {\bibfnamefont
  {F.}~\bibnamefont {Hatami}}, \bibinfo {author} {\bibfnamefont
  {W.}~\bibnamefont {Yao}}, \bibinfo {author} {\bibfnamefont {J.}~\bibnamefont
  {Vuckovic}}, \bibinfo {author} {\bibfnamefont {A.}~\bibnamefont {Majumdar}},
  \ and\ \bibinfo {author} {\bibfnamefont {X.}~\bibnamefont {Xu}},\ }\href
  {\doibase 10.1038/nature14290} {\bibfield  {journal} {\bibinfo  {journal}
  {Nature (London)}\ }\textbf {\bibinfo {volume} {520}},\ \bibinfo {pages} {69}
  (\bibinfo {year} {2015})}\BibitemShut {NoStop}%
\bibitem [{\citenamefont {Ye}\ \emph {et~al.}(2015)\citenamefont {Ye},
  \citenamefont {Wong}, \citenamefont {Lu}, \citenamefont {Ni}, \citenamefont
  {Zhu}, \citenamefont {Chen}, \citenamefont {Wang},\ and\ \citenamefont
  {Zhang}}]{Laser_Xiang}%
  \BibitemOpen
  \bibfield  {author} {\bibinfo {author} {\bibfnamefont {Y.}~\bibnamefont
  {Ye}}, \bibinfo {author} {\bibfnamefont {Z.~J.}\ \bibnamefont {Wong}},
  \bibinfo {author} {\bibfnamefont {X.}~\bibnamefont {Lu}}, \bibinfo {author}
  {\bibfnamefont {X.}~\bibnamefont {Ni}}, \bibinfo {author} {\bibfnamefont
  {H.}~\bibnamefont {Zhu}}, \bibinfo {author} {\bibfnamefont {X.}~\bibnamefont
  {Chen}}, \bibinfo {author} {\bibfnamefont {Y.}~\bibnamefont {Wang}}, \ and\
  \bibinfo {author} {\bibfnamefont {X.}~\bibnamefont {Zhang}},\ }\href
  {\doibase 10.1038/nphoton.2015.197} {\bibfield  {journal} {\bibinfo
  {journal} {Nat. Photonics}\ }\textbf {\bibinfo {volume} {9}},\ \bibinfo
  {pages} {733} (\bibinfo {year} {2015})}\BibitemShut {NoStop}%
\bibitem [{\citenamefont {Liu}\ \emph {et~al.}(2017)\citenamefont {Liu},
  \citenamefont {Clark}, \citenamefont {Fryett}, \citenamefont {Wu},
  \citenamefont {Zheng}, \citenamefont {Hatami}, \citenamefont {Xu},\ and\
  \citenamefont {Majumdar}}]{CHL_LED}%
  \BibitemOpen
  \bibfield  {author} {\bibinfo {author} {\bibfnamefont {C.-H.}\ \bibnamefont
  {Liu}}, \bibinfo {author} {\bibfnamefont {G.}~\bibnamefont {Clark}}, \bibinfo
  {author} {\bibfnamefont {T.}~\bibnamefont {Fryett}}, \bibinfo {author}
  {\bibfnamefont {S.}~\bibnamefont {Wu}}, \bibinfo {author} {\bibfnamefont
  {J.}~\bibnamefont {Zheng}}, \bibinfo {author} {\bibfnamefont
  {F.}~\bibnamefont {Hatami}}, \bibinfo {author} {\bibfnamefont
  {X.}~\bibnamefont {Xu}}, \ and\ \bibinfo {author} {\bibfnamefont
  {A.}~\bibnamefont {Majumdar}},\ }\href {\doibase
  10.1021/acs.nanolett.6b03801} {\bibfield  {journal} {\bibinfo  {journal}
  {Nano Letters}\ }\textbf {\bibinfo {volume} {17}},\ \bibinfo {pages} {200}
  (\bibinfo {year} {2017})},\ \bibinfo {note} {pMID: 27936763}\BibitemShut
  {NoStop}%
\bibitem [{\citenamefont {Liu}\ \emph {et~al.}(2014)\citenamefont {Liu},
  \citenamefont {Galfsky}, \citenamefont {Sun}, \citenamefont {Xia},
  \citenamefont {chen Lin}, \citenamefont {Lee}, \citenamefont {Kena-Cohen},\
  and\ \citenamefont {Menon}}]{Vinod_Menon_EP}%
  \BibitemOpen
  \bibfield  {author} {\bibinfo {author} {\bibfnamefont {X.}~\bibnamefont
  {Liu}}, \bibinfo {author} {\bibfnamefont {T.}~\bibnamefont {Galfsky}},
  \bibinfo {author} {\bibfnamefont {Z.}~\bibnamefont {Sun}}, \bibinfo {author}
  {\bibfnamefont {F.}~\bibnamefont {Xia}}, \bibinfo {author} {\bibfnamefont
  {E.}~\bibnamefont {chen Lin}}, \bibinfo {author} {\bibfnamefont {Y.-H.}\
  \bibnamefont {Lee}}, \bibinfo {author} {\bibfnamefont {S.}~\bibnamefont
  {Kena-Cohen}}, \ and\ \bibinfo {author} {\bibfnamefont {V.~M.}\ \bibnamefont
  {Menon}},\ }\href@noop {} {\ \textbf {\bibinfo {volume} {9}},\ \bibinfo
  {pages} {30} (\bibinfo {year} {2014})}\BibitemShut {NoStop}%
\bibitem [{\citenamefont {Dufferwiel}\ \emph {et~al.}(2015)\citenamefont
  {Dufferwiel}, \citenamefont {Schwarz}, \citenamefont {Withers}, \citenamefont
  {Trichet}, \citenamefont {Li}, \citenamefont {Sich}, \citenamefont
  {Pozo-Zamudio}, \citenamefont {Clark}, \citenamefont {Nalitov}, \citenamefont
  {Solnyshkov}, \citenamefont {Malpuech}, \citenamefont {Novoselov},
  \citenamefont {Smith}, \citenamefont {Skolnick}, \citenamefont
  {Krizhanovskii},\ and\ \citenamefont {Tartakovskii}}]{Sasha_EP}%
  \BibitemOpen
  \bibfield  {author} {\bibinfo {author} {\bibfnamefont {S.}~\bibnamefont
  {Dufferwiel}}, \bibinfo {author} {\bibfnamefont {S.}~\bibnamefont {Schwarz}},
  \bibinfo {author} {\bibfnamefont {F.}~\bibnamefont {Withers}}, \bibinfo
  {author} {\bibfnamefont {A.~A.~P.}\ \bibnamefont {Trichet}}, \bibinfo
  {author} {\bibfnamefont {F.}~\bibnamefont {Li}}, \bibinfo {author}
  {\bibfnamefont {M.}~\bibnamefont {Sich}}, \bibinfo {author} {\bibfnamefont
  {O.~D.}\ \bibnamefont {Pozo-Zamudio}}, \bibinfo {author} {\bibfnamefont
  {C.}~\bibnamefont {Clark}}, \bibinfo {author} {\bibfnamefont
  {A.}~\bibnamefont {Nalitov}}, \bibinfo {author} {\bibfnamefont {D.~D.}\
  \bibnamefont {Solnyshkov}}, \bibinfo {author} {\bibfnamefont
  {G.}~\bibnamefont {Malpuech}}, \bibinfo {author} {\bibfnamefont {K.~S.}\
  \bibnamefont {Novoselov}}, \bibinfo {author} {\bibfnamefont {J.~M.}\
  \bibnamefont {Smith}}, \bibinfo {author} {\bibfnamefont {M.~S.}\ \bibnamefont
  {Skolnick}}, \bibinfo {author} {\bibfnamefont {D.~N.}\ \bibnamefont
  {Krizhanovskii}}, \ and\ \bibinfo {author} {\bibfnamefont {A.~I.}\
  \bibnamefont {Tartakovskii}},\ }\href {\doibase 10.1038/ncomms9579}
  {\bibfield  {journal} {\bibinfo  {journal} {Nat. Commun.}\ }\textbf {\bibinfo
  {volume} {6}} (\bibinfo {year} {2015}),\ 10.1038/ncomms9579}\BibitemShut
  {NoStop}%
\bibitem [{\citenamefont {Wei}\ \emph {et~al.}(2017)\citenamefont {Wei},
  \citenamefont {Czaplewski}, \citenamefont {Lenferink}, \citenamefont
  {Stanev}, \citenamefont {Jung},\ and\ \citenamefont {Stern}}]{wei2017size}%
  \BibitemOpen
  \bibfield  {author} {\bibinfo {author} {\bibfnamefont {G.}~\bibnamefont
  {Wei}}, \bibinfo {author} {\bibfnamefont {D.~A.}\ \bibnamefont {Czaplewski}},
  \bibinfo {author} {\bibfnamefont {E.~J.}\ \bibnamefont {Lenferink}}, \bibinfo
  {author} {\bibfnamefont {T.~K.}\ \bibnamefont {Stanev}}, \bibinfo {author}
  {\bibfnamefont {I.~W.}\ \bibnamefont {Jung}}, \ and\ \bibinfo {author}
  {\bibfnamefont {N.~P.}\ \bibnamefont {Stern}},\ }\href {\doibase
  10.1038/s41598-017-03594-z} {\bibfield  {journal} {\bibinfo  {journal} {Sci.
  Rep.}\ }\textbf {\bibinfo {volume} {7}} (\bibinfo {year} {2017}),\
  10.1038/s41598-017-03594-z}\BibitemShut {NoStop}%
\bibitem [{\citenamefont {Fryett}\ \emph
  {et~al.}(2017{\natexlab{b}})\citenamefont {Fryett}, \citenamefont {Chen},
  \citenamefont {Whitehead}, \citenamefont {Peycke}, \citenamefont {Xu},\ and\
  \citenamefont {Majumdar}}]{fryett2017encapsulated}%
  \BibitemOpen
  \bibfield  {author} {\bibinfo {author} {\bibfnamefont {T.~K.}\ \bibnamefont
  {Fryett}}, \bibinfo {author} {\bibfnamefont {Y.}~\bibnamefont {Chen}},
  \bibinfo {author} {\bibfnamefont {J.}~\bibnamefont {Whitehead}}, \bibinfo
  {author} {\bibfnamefont {Z.~M.}\ \bibnamefont {Peycke}}, \bibinfo {author}
  {\bibfnamefont {X.}~\bibnamefont {Xu}}, \ and\ \bibinfo {author}
  {\bibfnamefont {A.}~\bibnamefont {Majumdar}},\ }\href@noop {} {\bibfield
  {journal} {\bibinfo  {journal} {arXiv preprint arXiv:1709.02032}\ } (\bibinfo
  {year} {2017}{\natexlab{b}})}\BibitemShut {NoStop}%
\bibitem [{\citenamefont {Sun}\ \emph {et~al.}(2017)\citenamefont {Sun},
  \citenamefont {Yoon}, \citenamefont {Steger}, \citenamefont {Liu},
  \citenamefont {Pfeiffer}, \citenamefont {West}, \citenamefont {Snoke},\ and\
  \citenamefont {Nelson}}]{Sun2017}%
  \BibitemOpen
  \bibfield  {author} {\bibinfo {author} {\bibfnamefont {Y.}~\bibnamefont
  {Sun}}, \bibinfo {author} {\bibfnamefont {Y.}~\bibnamefont {Yoon}}, \bibinfo
  {author} {\bibfnamefont {M.}~\bibnamefont {Steger}}, \bibinfo {author}
  {\bibfnamefont {G.}~\bibnamefont {Liu}}, \bibinfo {author} {\bibfnamefont
  {L.~N.}\ \bibnamefont {Pfeiffer}}, \bibinfo {author} {\bibfnamefont
  {K.}~\bibnamefont {West}}, \bibinfo {author} {\bibfnamefont {D.}~\bibnamefont
  {Snoke}}, \ and\ \bibinfo {author} {\bibfnamefont {K.~A.}\ \bibnamefont
  {Nelson}},\ }\href {\doibase 10.1038/nphys4148} {\bibfield  {journal}
  {\bibinfo  {journal} {Nat. Phys.}\ }\textbf {\bibinfo {volume} {13}},\
  \bibinfo {pages} {870 EP } (\bibinfo {year} {2017})},\ \bibinfo {note}
  {article}\BibitemShut {NoStop}%
\bibitem [{\citenamefont {Kavokin}\ \emph {et~al.}(2007)\citenamefont
  {Kavokin}, \citenamefont {Baumberg}, \citenamefont {Malpuech},\ and\
  \citenamefont {Laussy}}]{kavokin2007microcavities}%
  \BibitemOpen
  \bibfield  {author} {\bibinfo {author} {\bibfnamefont {A.}~\bibnamefont
  {Kavokin}}, \bibinfo {author} {\bibfnamefont {J.}~\bibnamefont {Baumberg}},
  \bibinfo {author} {\bibfnamefont {G.}~\bibnamefont {Malpuech}}, \ and\
  \bibinfo {author} {\bibfnamefont {F.}~\bibnamefont {Laussy}},\ }\href@noop {}
  {\emph {\bibinfo {title} {Microcavities, Series on Semiconductor Science and
  Technology}}}\ (\bibinfo  {publisher} {Oxford University Press New York},\
  \bibinfo {year} {2007})\BibitemShut {NoStop}%
\bibitem [{\citenamefont {Birnbaum}\ \emph {et~al.}(2005)\citenamefont
  {Birnbaum}, \citenamefont {Boca}, \citenamefont {Miller}, \citenamefont
  {Boozer}, \citenamefont {Northup},\ and\ \citenamefont
  {Kimble}}]{birnbaum2005photon}%
  \BibitemOpen
  \bibfield  {author} {\bibinfo {author} {\bibfnamefont {K.~M.}\ \bibnamefont
  {Birnbaum}}, \bibinfo {author} {\bibfnamefont {A.}~\bibnamefont {Boca}},
  \bibinfo {author} {\bibfnamefont {R.}~\bibnamefont {Miller}}, \bibinfo
  {author} {\bibfnamefont {A.~D.}\ \bibnamefont {Boozer}}, \bibinfo {author}
  {\bibfnamefont {T.~E.}\ \bibnamefont {Northup}}, \ and\ \bibinfo {author}
  {\bibfnamefont {H.~J.}\ \bibnamefont {Kimble}},\ }\href {\doibase
  10.1038/nature03804} {\bibfield  {journal} {\bibinfo  {journal} {Nature
  (London)}\ }\textbf {\bibinfo {volume} {436}},\ \bibinfo {pages} {87}
  (\bibinfo {year} {2005})}\BibitemShut {NoStop}%
\bibitem [{\citenamefont {Bamba}\ \emph {et~al.}(2011)\citenamefont {Bamba},
  \citenamefont {Imamo\ifmmode~\breve{g}\else \u{g}\fi{}lu}, \citenamefont
  {Carusotto},\ and\ \citenamefont {Ciuti}}]{bamba2011origin}%
  \BibitemOpen
  \bibfield  {author} {\bibinfo {author} {\bibfnamefont {M.}~\bibnamefont
  {Bamba}}, \bibinfo {author} {\bibfnamefont {A.}~\bibnamefont
  {Imamo\ifmmode~\breve{g}\else \u{g}\fi{}lu}}, \bibinfo {author}
  {\bibfnamefont {I.}~\bibnamefont {Carusotto}}, \ and\ \bibinfo {author}
  {\bibfnamefont {C.}~\bibnamefont {Ciuti}},\ }\href {\doibase
  10.1103/PhysRevA.83.021802} {\bibfield  {journal} {\bibinfo  {journal} {Phys.
  Rev. A}\ }\textbf {\bibinfo {volume} {83}},\ \bibinfo {pages} {021802}
  (\bibinfo {year} {2011})}\BibitemShut {NoStop}%
\bibitem [{\citenamefont {Liew}\ and\ \citenamefont
  {Savona}(2010)}]{liew2010single}%
  \BibitemOpen
  \bibfield  {author} {\bibinfo {author} {\bibfnamefont {T.~C.~H.}\
  \bibnamefont {Liew}}\ and\ \bibinfo {author} {\bibfnamefont {V.}~\bibnamefont
  {Savona}},\ }\href {\doibase 10.1103/PhysRevLett.104.183601} {\bibfield
  {journal} {\bibinfo  {journal} {Phys. Rev. Lett.}\ }\textbf {\bibinfo
  {volume} {104}},\ \bibinfo {pages} {183601} (\bibinfo {year}
  {2010})}\BibitemShut {NoStop}%
\bibitem [{\citenamefont {Wang}\ \emph {et~al.}(2017)\citenamefont {Wang},
  \citenamefont {Zhan}, \citenamefont {Xu}, \citenamefont {Chen}, \citenamefont
  {You}, \citenamefont {Majumdar},\ and\ \citenamefont
  {Jiang}}]{wang2017quantum}%
  \BibitemOpen
  \bibfield  {author} {\bibinfo {author} {\bibfnamefont {H.-X.}\ \bibnamefont
  {Wang}}, \bibinfo {author} {\bibfnamefont {A.}~\bibnamefont {Zhan}}, \bibinfo
  {author} {\bibfnamefont {Y.-D.}\ \bibnamefont {Xu}}, \bibinfo {author}
  {\bibfnamefont {H.-Y.}\ \bibnamefont {Chen}}, \bibinfo {author}
  {\bibfnamefont {W.-L.}\ \bibnamefont {You}}, \bibinfo {author} {\bibfnamefont
  {A.}~\bibnamefont {Majumdar}}, \ and\ \bibinfo {author} {\bibfnamefont
  {J.-H.}\ \bibnamefont {Jiang}},\ }\href {\doibase 10.1088/1361-648X/aa8933}
  {\bibfield  {journal} {\bibinfo  {journal} {J. Phys. Condens. Matter}\
  }\textbf {\bibinfo {volume} {29}},\ \bibinfo {pages} {445703} (\bibinfo
  {year} {2017})}\BibitemShut {NoStop}%
\end{thebibliography}%

\end{document}